\documentclass[10pt]{article}
\usepackage[T1]{fontenc}
\usepackage{graphicx}
\usepackage{amssymb}
\usepackage{amsfonts}
\usepackage{amsmath}
\usepackage{listings}
\usepackage{enumerate}
\usepackage{tikz,tikz-cd}
\usepackage{todonotes}
\usepackage{multirow}
\usepackage[hyphens]{url}
\usepackage{hyperref}
\usepackage{eurosym}
\usepackage{amsfonts}
\usepackage{color}
\usepackage{amsopn}
\usepackage{latexsym}
\usepackage{float}
\usepackage{bm}
\usepackage{xcolor}
\usepackage[utf8]{inputenc}
\usepackage{hhline} 
\usepackage{verbatim}
\usepackage[hyphenbreaks]{breakurl}

\usepackage{titlesec}

\titleformat{\paragraph}[runin]
  {\normalfont\itshape}
  {\theparagraph}
  {}
  {}
  []

\newtheorem{theorem}{Theorem}[section] 
 
\newtheorem{definition}[theorem]{Definition}

\newcommand{\mz}{\mathfrak z }

\newcommand{\br}{[\,\cdot\,,\cdot\,]}

\newcommand{\mcD}{\mathcal D}

\newcommand{\R}{\mathbb R}

\newcommand{\N}{\mathbb N}
\newcommand{\Z}{\mathbb Z}
\newcommand{\C}{\mathbb C}

\newcommand{\K}{\mathbb K}

\DeclareMathOperator{\End}{End}

\DeclareMathOperator{\Der}{Der}

\DeclareMathOperator{\ad}{ad}

\numberwithin{equation}{section}

\newcommand{\nc}{\newcommand}
\nc{\Fin}{\operatorname{Fin}}
\nc{\Iso}{\operatorname{Iso}}
\nc{\Id}{\operatorname{Id}}
 \nc{\iso}{\mathfrak{iso}}
 \nc{\sso}{\mathfrak{so}}
\nc{\Ad}{\operatorname{Ad}} 
\nc{\Sym}{\mathrm{Sym}}
 \nc{\pr}{\operatorname{pr}} 
 \nc{\Dera}{\operatorname{Dera}} 
 \nc{\Auto}{\operatorname{Auto}}
 \nc{\noi}{\noindent}
 \nc{\SO}{\operatorname{\rm SO}}
\nc{\aff}{\mathfrak{aff}}
\nc{\hyp}{\mathfrak{hyp}}
\nc{\heis}{\mathfrak{heis}}

\definecolor{keywordcolor}{rgb}{0.7, 0.1, 0.1}   
\definecolor{tacticcolor}{rgb}{0.0, 0.1, 0.6}    
\definecolor{commentcolor}{rgb}{0.4, 0.4, 0.4}   
\definecolor{symbolcolor}{rgb}{0.0, 0.1, 0.6}    
\definecolor{sortcolor}{rgb}{0.1, 0.5, 0.1}      
\definecolor{attributecolor}{rgb}{0.7, 0.1, 0.1} 

\title{Formalizing a classification theorem for
low-dimensional solvable Lie algebras in Lean}

\author{Viviana del Barco\and Gustavo Infanti\and Exequiel Rivas \and Paul Schwahn}

\newcommand{\keywords}[1]{\par\noindent\textbf{Keywords:} #1}

\begin{document}

\maketitle              
\begin{abstract}
We present a formalization, in the theorem prover \emph{Lean}, of the classification of solvable Lie algebras of dimension at most three over arbitrary fields. Lie algebras are algebraic objects which encode infinitesimal symmetries, and as such ubiquitous in geometry and physics. Our work involves explicit calculations on the level of the underlying vector spaces and provides a use case for the linear algebra and Lie theory routines in Lean's mathematical library \emph{mathlib}. Along the way we formalize results about Lie algebras, define the semidirect product within this setting and add API for bases of vector spaces. In a wider context, this project aims to provide a complete mechanization of a classification theorem, covering both the statement and its full formal proof, and contribute to the development and broader adoption of such results in formalized mathematics.
\end{abstract}

\medskip

\section{Introduction}

In recent years, the formalization of mathematics has attracted
growing interest among mathematicians. 
Formalization tools are of great value not only for reducing
human error and possible oversights, but also for making all assumptions and
dependencies fully explicit, providing interesting insights into the
structure of mathematical arguments.
One of these tools, the Lean proof assistant~\cite{lean4}, has gained widespread adoption thanks to its expressive
foundations, a predicative calculus of inductive constructions, and its powerful elaborator.
Lean is complemented by mathlib~\cite{mathlib4}, a
community-maintained library that includes formalization of a broad spectrum of
mathematical fields, including group theory, commutative algebra,
linear algebra, differential geometry, and measure theory.
This article is about the formalization of a classification theorem in a field
of mathematics already present in mathlib: Lie algebras.

The origins of Lie algebras and Lie groups date back to the work of Sophus Lie in late 18th century, who introduced them in order to study continuous symmetries of differential equations. Lie groups can be understood as continuous groups of transformations, while Lie algebras are sets of "infinitesimal" symmetries.
Today, Lie groups and Lie algebras are widely used and have shaped the fields of geometry and algebra. Attempts at classifications of Lie algebras began with Killing, and independently Cartan, followed by Bianchi.

In mathematics, a \emph{full classification} of a collection of objects is understood to be a non-redundant, exhaustive list of equivalence classes of those objects, for an appropriate notion of \emph{equivalence} or \emph{isomorphism} -- each of the objects in the collection  belongs to precisely one of the equivalence classes. 
One might want not only to \emph{classify} mathematical objects, but also to \emph{characterize} the invidual cases in the classification -- either through giving some invariant that separates the cases, or by stating equivalent conditions for each case.

This general perspective naturally applies to Lie algebras. One may reasonably ask to classify the Lie algebras of a certain type (simple, solvable etc.) or dimension, up to isomorphism. Each class should be represented by a particular Lie algebra, given either in terms of generators and Lie bracket relations or through another construction (like a semidirect product).

For the class of finite-dimensional complex simple Lie algebras, Killing provided a classification whose list was not complete. The complete classification is due to Cartan a few years later. Levi's theorem states that arbitrary complex Lie  algebras are  built from simple Lie algebras and solvable Lie algebras. The classification of the latter would thus lead to a full classification of  complex Lie algebras. However, due to the many degrees of freedom that solvable Lie algebras have, such a classification is hopeless in full generality -- hence authors focus on specific subfamilies. 

One such subclass is the family of nilpotent Lie algebras of dimension 6. Between 1958 and 1983, several attempts were made to classify these over various fields:  Morozov \cite{Mo58} (in characteristic zero), Shedler \cite{Sh64} (any field), Vergne \cite{Ve66} (over $\C$), Skjelbread and Sun \cite{SS78} (over $\R$), Bech and Kolman \cite{BK81} (over $\R$). Many of these works are independent and use different methods -- however, in some cases, the lists are not compatible. It is Nielsen \cite{Nielsen} who compared these tables and produced a first complete list of 6-dimensional nilpotent Lie algebras over the reals. 

Similar challenges arose in dimension 7, as documented by Gong \cite{Gong98}. It is not unusual in the literature to find multiple classification papers on the same class of algebras, differing only in minor details, and coexisting until a consensus is reached around a \emph{definitive} classification. Even then, in large classifications, cases can be overlooked and only noticed later (if at all).

This is why we believe that formalization of classifications, for example of Lie algebras, in a theorem prover such as Lean will constitute a significant contribution to the mathematical community. Formalized classifications will remove all possible doubts of completeness and correctness of the provided classifications lists. Classifications will not depend on accepted lists, but on rigorous formalized proofs.

A number of mathematical classification results have already been formalized in theorem provers. 
Focusing on Lean, the well known classification or \emph{structure theorem} for finite abelian groups \cite{finiteabelian}
and a classification theorem for finite Z-groups \cite{zgroup} are already formalized in mathlib. Another notable recent project formalizes the classification of groups of order $pq$ where $p$, $q$ are primes \cite{pqgroups}.

The current paper contributes to the formalization in Lean of classification results
about Lie algebras. Concretely, we formalize a full classification theorem for solvable Lie algebras of dimension $\leq 3$ over arbitrary fields. Moreover, we provide characterizations for  each of the cases through an
equivalent condition. Even though the mathematical correctness of this classification is not in doubt, its formalization -- including all necessary supporting lemmas -- is an important step for further formal work in Lie theory. Such a formalization requires the basic concepts of Lie theory to be formalized in Lean and included in the community maintained Lean library \emph{mathlib}. This task was started  by Nash \cite{Nash:2022}, who has an ongoing program to formalize Cartan's classification of finite-dimensional complex simple Lie algebras. 

More specifically, in the context of Levi's theorem -- and given that any Lie algebra of dimension  at most 3 is either solvable or simple -- our formalization is in some sense complementary to Nash's work. Our formalization follows the most recent and widely accepted classification of 3-dimensional solvable Lie algebras due to de Graaf \cite{DeGraaf:2005}.

More precisely, we formalize the following result:
\begin{theorem}\label{thm:main}
    Let $L$ be a solvable Lie algebra of dimension $\leq 3$ over a field $\K$. 
    \begin{enumerate}[\upshape(i)]
        \item If $\dim L=1$, then $L$ is isomorphic to the abelian Lie algebra $\K$.
        \item If $\dim L=2$, then $L$ is isomorphic to the abelian Lie algebra $\K^2$ or to $\aff(\K)$.       
        \item If $\dim L=3$, then $L$ is isomorphic to one of the following Lie algebras: $\K^3$, $\heis_3(\K)$, $\aff(\K)\oplus \K$,  $\hyp_3(\K)$, $L_\alpha$ for some $\alpha\in\K^{\times}$, or $M_{\delta}$ for some $\delta\in \K^{\times}/(\K^{\times})^2$, where $(\K^{\times})^2$ denotes the subgroup of squares of $\K^\times$. 
    \end{enumerate}
    The Lie algebras listed above are pairwise non-isomorphic.
\end{theorem}
The Lie algebras appearing here are described in detail in Table~\ref{tab:not} and the basics on Lie algebras can be found in Section \ref{sec:basic}.

The formalization of this theorem requires several constructions in Lean: first, to introduce all the Lie algebra structures mentioned there ($\K^n, \aff(\K), \heis_3(\K)$, etc.) as an \lstinline{instance} of a typeclass in Lean. We briefly explain the procedure here; for more details, see Section \ref{sec:instances}. We model each Lie algebra above on $\K^n$ (the $n$-th cartesian power of $\K$) and explicitly provide the Lie bracket for each one. Part of declaring the Lie algebra instance is to prove in Lean that this bracket indeed satisfies the Lie algebra axioms. After this is done, we show in Section \ref{sec:classif} that given a solvable Lie algebra of dimension $\leq 3$, there exists a Lie algebra isomorphism to one of the Lie algebras listed above. This is done by using bases of the Lie algebras (as vector spaces). The last (but not least!) part is to show that the defined instances are pairwise non-isomorphic. This is done by formalizing \emph{invariants} of Lie algebras, that is, properties that remain invariant under isomorphisms. For instance, isomorphic Lie algebras are of the same dimension, just as vector spaces. In addition, we show in Lean that they must also have the same dimension of the commutator ideal. As a consequence, an abelian and a non-abelian Lie algebra cannot be isomorphic; in particular, the two Lie algebras above in (ii) are not. More involved invariants were formalized and used to distinguish the Lie algebras in (iii). A detailed description of this process is presented in Section \ref{sec:classif}, and the full Lean code is available at \cite{dBORS}.

Section \ref{sec:contr} contains the  auxiliary material formalized, consisting on many basic results on vector spaces and linear maps—along with new results and API on top of the already formalized linear algebra and Lie theory in mathlib—with the goal of eventually contributing them back to the library. In addition, we provide results on (solvable) Lie algebras that we plan to contribute to mathlib. A few of them are particularly remarkable: 
\begin{itemize}
    \item the formalization of the semidirect product of Lie algebras,
    \item writing the Lie algebras in Theorem~\ref{thm:main} as semidirect products of smaller pieces, in particular showing that they are all \emph{almost abelian} (see Section \ref{sec:alab} for the precise definition),
    \item the proof that a Lie algebra is solvable if a quotient by some solvable ideal is solvable. 
\end{itemize}

One of the frequent lessons in formalization of mathematics is that writing good definitions and theorem statements is just as hard as writing proofs. When formalizing a classification, care must be taken when selecting the representatives for the isomorphism classes and when stating the theorems about isomorphisms, non-isomorphisms, and characterizations of the cases. Since our project aspires to all of this, it may serve as a reference point for designing further classification projects in mathematics.

Development information on the project is given in Section \ref{sec:dev}, including the link to our formalization repository. Finally, in Section \ref{sec:future}, we comment on future work we plan to approach.
\medskip

\noindent {\bf Acknowledgments} V.~del Barco is supported by FAPESP grant 2023/15089-9. G.~Infanti was supported by undergraduate fellowship PICME-CNPQ. P.~Schwahn is supported by the FAPESP grant 2024/08127-4. E.~Rivas was supported by the Estonian Research Council grant no.~PSG749.

\section{Background}\label{sec:basic}

\subsection{Mathematical preliminaries}

We begin with the mathematical definition of a Lie algebra as it is often found in textbooks.

\begin{definition}\label{def:liealg}
    A vector space $L$ over a field $\K$ together with an operation $\br:L\times L\to L$ is called a \emph{Lie algebra} over $\K$ if the following axioms are satisfied:
    \begin{enumerate}[\upshape(i)]
        \item $\br:L\times L\to L$ is bilinear,
        \item $[x,x]=0$ for all $x\in L$,
        \item $[x,[y,z]]+[y,[z,x]]+[z,[x,y]]=0$ for all $x,y,z\in L$.
    \end{enumerate}
\end{definition}

The operation $\br:L\times L\to L$ is called the \emph{Lie bracket} of $L$. The second condition is called skew-symmetry of the bracket since it implies $[x,y]=-[y,x]$ for all $x,y\in L$. The last condition is known as the \emph{Jacobi identity}.

We note that it is possible to relax some assumptions in the above definition. For example, one may merely assume that $\K$ is a commutative ring instead of a field, as is done in mathlib (see Sect.~\ref{sec:prelimlean}). However, for the purpose of this paper as well as for most applications (in relations to Lie groups or algebraic groups) it suffices to consider Lie algebras over fields.

\paragraph{Abelian Lie algebras.}
The simplest possible Lie bracket is of course the trivial one, i.e.~$[x,y]=0$ for all $x,y\in L$. If this is the case, the Lie algebra $(L,\br)$ is called \emph{abelian}.

\paragraph{Lie subalgebras.}
A vector subspace $L'\subseteq L$ is called a \emph{Lie subalgebra} of $(L,\br)$ if it is closed under the Lie bracket of $L$, that is $[x,y]\in L'$ for all $x,y\in L'$.

\paragraph{Maps between Lie algebras.}
Let $(L_1,\br_1)$ and $(L_2,\br_2)$ be Lie algebras over a field $\K$. If  $f: L_1\to L_2$ is a linear map and preserves the Lie bracket, i.e.
\[f([x,y]_1)=[f(x),f(y)]_2\qquad\forall x,y\in L_1,\]
it is called a \emph{homomorphism} of Lie algebras. As always, an invertible homomorphism is called an \emph{isomorphism}. This will be the notion of equivalence for all classification theorems about Lie algebras.

A linear map $D: L\to L$ from a Lie algebra $(L,\br)$ to itself is called a \emph{derivation} if
\[D([x,y])=[D(x),y]+[x,D(y)]\qquad\forall x,y\in L.\]
The set of all derivations of $L$ is denoted by $\Der(L)$, which is itself a Lie algebra with Lie bracket defined as $[A,B]=AB-BA$, for $A,B\in \Der(L)$.

\paragraph{Ideals.}
If a vector subspace $I\subseteq L$ is \emph{absorbing}, that is $[x,y]\in I$ for all $x\in L$ and $y\in I$, we say that $I$ is an \emph{ideal} of $L$. Ideals are to Lie algebras what normal subgroups are to groups. For example, the kernel of a Lie algebra homomorphism is always an ideal. Moreover, if $I\subset L$ is an ideal, then the Lie bracket descends to the (vector space) quotient $L/I$ and turns it into a Lie algebra.

The \emph{center} of $L$ is defined as $\mz(L)=\{x\in L\,|\,[x,y]=0\ \forall y\in L\}$ and is an ideal of $L$. Clearly, $L$ is abelian if and only if it coincides with its center. A non-abelian Lie algebra $L$ which does not have any ideals other than $\{0\}$ and $L$ itself is called \emph{simple}.

\paragraph{Derived series.}
For ideals $I,J\subset L$ we declare $[I,J]\subseteq I\cap J$ to be the subspace of $L$ spanned by the set $\{[x,y]\,|\,x\in I,\ y\in J\}$. The ideal $[L,L]$ is called the \emph{commutator} or \emph{derived subalgebra} of $L$, and it will be a valuable tool in the proof of the classification theorem. The \emph{derived series} of a Lie algebra is defined as the series of repeated commutators,
\[\mcD^0(L):=L,\qquad\mcD^k(L):=[\mcD^{k-1}(L),\mcD^{k-1}(L)]\quad\text{for }k\geq 1.\]
Clearly, $\mcD^0(L)\supseteq \mcD^1(L)\supseteq \cdots\supseteq \mcD^{k}(L)\supseteq \mcD^{k+1}(L)\supseteq \cdots$. A Lie algebra is called \emph{solvable} if this series terminates
eventually, i.e.~$\mcD^k(L)=0$ for some $k\in \N$. In particular, any abelian Lie algebra is solvable.

It follows from the definitions above that a Lie algebra cannot be both solvable and simple. These two conditions are in a sense complementary for characteristic zero: any Lie algebra can be decomposed as the semidirect product of a solvable and a simple one, as a consequence of Levi's theorem \cite{Jac}. In dimension at most three, it is well known that a Lie algebra is either simple or solvable, independently of $\K$. 

\subsection{Realization in Lean}
\label{sec:prelimlean}

In what follows, we summarize the existing implementation of the most relevant concepts above in the Lean library \emph{mathlib}. To endow mathematical objects with structure, mathlib makes heavy use of the \emph{typeclass} system on top of Lean's type-theoretic foundations. Roughly, a mathematical object or set is reinterpreted as a type \lstinline{T : Type*}, and additional structure is encoded as an instance of a typeclass depending on \lstinline{T}. These instances are usually not named and are automatically inferred when needed. For example, a field $(\K,+,\cdot)$ is formalized as a type \lstinline{K : Type*} together with an instance of the typeclass \lstinline{Field K}, which bundles the addition and multiplication together with the proofs of the field axioms.

A \emph{vector space (or module)} $V$ of $\K$ is then understood as a type \lstinline{V} with instances of the typeclasses \lstinline{AddCommGroup V} and \lstinline{Module K V}. On the other hand, it has proved useful to define a \emph{subspace (or submodule)} of $V$ not as a standalone type (since there is no subset relation between types), but as a \emph{structure} \lstinline{Submodule K V}, which is essentially a bundle of data, consisting of a set of elements of \lstinline{V} (currently implemented as a \emph{predicate} on \lstinline{V}) together with the proofs that it contains zero and is closed under addition and scalar multiplication.
\begin{lstlisting}
structure Submodule (K : Type u) (V : Type v)
        [Semiring K] [AddCommMonoid V] [Module K V] : Type v where
    carrier : Set V
    add_mem' {a b : V} : a ∈ self.carrier → b ∈ self.carrier → a + b ∈ self.carrier
    zero_mem' : 0 ∈ self.carrier
    smul_mem' (c : K) {x : V} : x ∈ self.carrier → c • x ∈ self.carrier
\end{lstlisting}
Of course any subspace of $V$ is a vector space in its own right -- this leads to the use of \emph{type coercions}, which we will briefly discuss at the beginning of Section~\ref{sec:contr}.

A formalization of the \emph{dimension} of a finite dimensional vector space (that is, the rank as a module) is \lstinline{Module.finrank}, which is a natural number.

Let us turn to some implementation details that are specific to the context of Lie algebras. In mathlib there is a typeclass encoding bracket operations $\br$ without any extra conditions (and motivated from modules of Lie algebras, slightly more general than what is stated above):
\begin{lstlisting}
class Bracket (L M : Type*) where
    bracket : L → M → M
\end{lstlisting}
In mathlib, a \emph{Lie ring} is an additive group with a compatible bi-additive bracket that is skew-symmetric and satisfies the Jacobi identity, and a Lie algebra is a Lie ring \lstinline{L} which is also a module over some commutative ring \lstinline{K} such that the bracket becomes bilinear.
\begin{lstlisting}
class LieRing (L : Type*) extends AddCommGroup L, Bracket L L where [...]
class LieAlgebra (K L : Type*) [CommRing K] [LieRing L] extends Module K L where [...]
\end{lstlisting}

Just as a submodule of a module is a bundled structure in mathlib, the same goes for Lie subalgebras and ideals of a Lie algebra $L$, which are implemented as \lstinline{LieSubalgebra K L} and \lstinline{LieIdeal K L}, respectively.

The current paper deals with particular Lie algebras, namely, the solvable ones. In Lean, the solvability property is given by a typeclass:
\begin{lstlisting}
class LieAlgebra.IsSolvable (L : Type*) [LieRing L] : Prop where
    solvable_int : ∃ k, derivedSeries ℤ L k = ⊥
\end{lstlisting}
Note that every Lie ring is a Lie algebra over $\Z$, as instantiated in \lstinline{LieRing.instLieAlgebra}, just like any abelian group is a $\Z$-module. For a Lie algebra over another field, this definition of solvability coincides with the usual one given above, as formalized in the following mathlib theorem:
\begin{lstlisting}    
theorem LieAlgebra.isSolvable_iff  (K L : Type*) [CommRing K] [LieRing L] [LieAlgebra K L] :
    IsSolvable L ↔ ∃ (k : ℕ), derivedSeries K L k = ⊥
\end{lstlisting}
There is a similar typeclass \lstinline{IsLieAbelian} for the property of being abelian, and the fact that an abelian Lie algebra is solvable manifests as an instance
\begin{lstlisting}
instance LieAlgebra.ofAbelianIsSolvable (L : Type*) [LieRing L] [IsLieAbelian L] :
    IsSolvable L
\end{lstlisting}
Finally, homomorphisms between Lie algebras $L$ and $L'$ are formalized by the structure \lstinline{LieHom K L L'}, which may be abbreviated as \lstinline{L →ₗ⁅K⁆ L'}. This structure is then extended to Lie algebra isomorphisms as \lstinline{LieEquiv K L L'}, which essentially bundle the underlying homomorphism with its inverse, and is also notated as \lstinline{L ≃ₗ⁅K⁆ L'}.

\section{The Lie algebra instances}\label{sec:instances}

In this section we explain how we use the mathlib library to instantiate each Lie algebra appearing in Theorem \ref{thm:main}. Even though the theorem requires $\K$ to be a field, the definitions below make also sense for any commutative ring. Declare the following variables:
\begin{lstlisting}
variable (K : Type*) [CommRing K]
\end{lstlisting}

As introduced in Definition \ref{def:liealg}, Lie algebras are finite-dimensional vector spaces together with a skew-symmetric bilinear map satisfying the Jacobi condition. Since all vector spaces of the same dimension over a field $\K$ are isomorphic, we may model any $n$-dimensional Lie algebra over a field $\K$ on the vector space $\K^n$. This could be represented in at least two ways: either as the iterated cartesian product $\K\times \K\times \cdots\times \K$ or as the set $\K^{\{0,\ldots,n-1\}}$ of all possible maps from the finite set of $n$ elements to $\K$.

Since the product is a binary operation and one would have to define \lstinline{K × (K × (K × ... × K))} recursively, the second option \lstinline{Fin n → K} turns out to be more practical and scalable. Tuples of elements can then be constructed using vector notation, e.g.~for \lstinline{a b c : K} we have \lstinline{![a, b, c] : Fin 3 → K}.

It is worth mentioning that the type \lstinline{Fin n → K} already comes with a \lstinline{Module K} instance in mathlib, derived from the pointwise structure on $\K$. We may now extend this to a Lie algebra structure. 

\paragraph{The Lie algebra $\aff(\K)$.}
As an example, we describe in detail the set up needed to define the Lie algebra of affinities $\aff(\K)$, the (up to isomorphism) unique non-abelian Lie algebra in dimension two. First, we define \lstinline{Affine K} as a type synonym in order to avoid conflicts in the instance inference system.
\begin{lstlisting}
namespace LieAlgebra.Dim2
def Affine := Fin 2 → K

instance : LieRing (Affine K) := {
    (inferInstance : AddCommGroup (Fin 2 → K)) with
    bracket := fun l r ↦ ![0, l 0 * r 1 - r 0 * l 1]
    [...] }

instance : LieAlgebra K (Affine K) := {
    (inferInstance : Module K (Fin 2 → K)) with
    lie_smul := [...] }
\end{lstlisting}
The keyword \lstinline{inferInstance} helps carrying over the existing instances from \lstinline{Fin 2 → K}. The omitted parts \lstinline{[...]} contain the proofs that the bracket is bi-additive, skew-symmetric, Jacobian, and in the last instance compatible with scalar multiplication.

In this way, most of the Lie algebras appearing in Theorem~\ref{thm:main} are defined ad hoc. These can also be obtained as more general dimension-independent constructions that we implemented in Lean.

\paragraph{Families in dimension three.}

Particular attention should be paid to the two one-parameter families $L_\alpha$ and $M_\delta$ appearing in Theorem \ref{thm:main}. These can be more elegantly subsumed as special cases of a 2-parameter family of Lie algebras, as is done in \cite{DeGraaf:2005}. We may easily have the type synonym and Lie ring instance depend on the explicit parameters $\alpha,\beta$ showing up in the Lie bracket:\pagebreak
\begin{lstlisting}
namespace LieAlgebra.Dim3
def Family (_ _ : K) := Fin 3 → K

instance (α β : K) : LieRing (Family K α β) := {
    (inferInstance : AddCommGroup (Fin 3 → K)) with
    bracket := fun l r ↦ ![0, (l 0 * r 2 - l 2 * r 0) * α,
        (l 0 * r 2 - l 2 * r 0) * β + l 0 * r 1 - l 1 * r 0]  
    [...] }
\end{lstlisting}
\begin{lstlisting}
instance (α β : K) : LieAlgebra K (Family K α β) := [...]
\end{lstlisting}
We then set $L_\alpha:=F_{\alpha,0}$ and $M_\alpha:=F_{\alpha,1}$. Later we show that if $\alpha=\gamma^2\alpha'$ for some $\gamma\in\K^\times$, then $M_{\alpha}\cong M_{\alpha'}$, thus the isomorphism class $M_{[\alpha]}$, $[\alpha]\in\K^\times/(\K^\times)^2$, is well-defined.

\paragraph{Abelian Lie algebras.}
The instances for the abelian Lie algebras were constructed a bit differently. We would like to view an arbitrary vector space as an abelian Lie algebra (i.e.~with trivial Lie bracket). Clearly it is inadvisible to directly define such a Lie bracket for any type \lstinline{V} with a \lstinline{Module K V} instance, since this will lead to conflicts in the instance inference system. Thus we make use of a type synonym
\begin{lstlisting}
namespace LieAlgebra
variable (K : Type*) [Field K] (V : Type*) [AddCommGroup V] [Module K V]

def mkAbelian := V
\end{lstlisting}
where we set the bracket of any two elements to be zero:
\begin{lstlisting}
instance : Bracket (mkAbelian K V) (mkAbelian K V) where
    bracket := fun _ _ ↦ (0 : V)
\end{lstlisting}
For the Lie ring and Lie algebra structures, we proceed as above. In this way, for any $\K$-module $V$, \lstinline{mkAbelian K V} represents the Lie algebra $V$ with trivial Lie bracket. In order to work with the abelian Lie algebras of dimension $\leq 3$, we set
\begin{lstlisting}
abbrev Dim1.Abelian := K
abbrev Dim2.Abelian := mkAbelian K (Fin 2 → K)    
abbrev Dim3.Abelian := mkAbelian K (Fin 3 → K)    
\end{lstlisting}
(conveniently, \lstinline{K} already inherits an abelian Lie algebra structure from its (commutative) algebra structure via the mathlib instance \lstinline{LieAlgebra.ofAssociativeAlgebra}).

\paragraph{Summary.}
For the reader's convenience, in Table~\ref{tab:not} we list the labels we used in the Lean code for each Lie algebra mentioned in Theorem \ref{thm:main}. In the last line of the table, $[\alpha]$ denotes the class of $\alpha\in \K^\times$ in the quotient $\K^\times/(\K^\times)^2$. We also list the non-zero brackets for a suitable basis $(b_i)$ of the Lie algebra.

\begin{table}[h]
\centering
\renewcommand\arraystretch{1.2}
\begin{tabular}{|c|c|c|c|}
\hline
\textbf{Dim} & \textbf{Notation} & \textbf{Lean name} & \textbf{Non-zero brackets}\\
\hline\hline
1 & $\K$ & \lstinline|K| & -- \\\hline
\multirow{2}{*}{2} & $\K^2$ & \lstinline|Dim2.Abelian K| & --\\
& $\aff(\K)$ & \lstinline|Dim2.Affine K| & $[b_0,b_1]=b_1$\\\hline
\multirow{6}{*}{3}& $\K^3$ & \lstinline|Dim3.Abelian K| & --\\
& $\heis_3(\K)$&\lstinline|Dim3.Heisenberg K| & $[b_1, b_2]=b_0$\ \\ 
& $\aff(\K)\oplus \K$&\lstinline|Dim3.AffinePlusAbelian K|& $[b_1,b_2]=b_1$\\
& $\hyp_3(\K)$&\lstinline|Dim3.Hyperbolic K|& $[b_0,b_1]=b_1,\,[b_0,b_2]=b_2$\\
& $F_{\alpha,\beta},\ \alpha\in \K^\times,\ \beta\in\K$ & \lstinline| Dim3.Family K α β, α ≠ 0| & $[b_0, b_1]=b_2,\,[b_0,b_2]=\alpha b_1+\beta b_2$\\
& $L_\alpha,\ \alpha\in \K^\times$ & \lstinline|Dim3.Family K α 0, α ≠ 0| & see above\\
& $M_{[\alpha]},\ \alpha \in \K^\times$ & \lstinline|Dim3.Family K α 1, α ≠ 0| & see above\\
\hline
\end{tabular}
\vspace{5pt}
\caption{Notation for low-dimensional Lie algebras.}
\label{tab:not}
\end{table}

\section{The classification theorem}
\label{sec:classif}

This section presents the required steps to formalize the proof of Theorem \ref{thm:main}. The formalization is divided according to the (finite) dimension of the Lie algebras. As seen in Section \ref{sec:basic}, if two Lie algebras are isomorphic, then they are isomorphic as vector spaces and thus have the same dimension. This is already formalized in mathlib: the dimension of vector spaces is preserved under linear isomorphism by \lstinline{LinearEquiv.finrank_eq}, and any isomorphism of Lie algebras is a linear isomorphism by \lstinline{LieEquiv.coe_toLinearEquiv}.

In order to get the full proof of the theorem, we first focus on showing that the Lie algebras in Section \ref{sec:instances} are a set of representatives for the isomorphism classes in the classification -- that is, every solvable Lie algebra of dimension $\leq 3$ is isomorphic to one of them. In a second stage, we prove that these representatives are pairwise non-equivalent. These two stages are described separately in the following two subsections. 

\subsection{Establishing isomorphisms with representatives}

For each $n=1,2,3$, we show that given a solvable Lie algebra of dimension $n$, there is an isomorphism to one of the Lie algebras from Table~\ref{tab:not}. The precise statements are as follows (from now on $\K$ is a field).

\begin{lstlisting}
variable {K L : Type*} [Field K] [LieRing L] [LieAlgebra K L]

theorem Dim1.classification (h : Module.finrank K L = 1) :
    Nonempty (L ≃ₗ⁅K⁆ K) 
    
theorem Dim2.classification (h : Module.finrank K L = 2) : 
    Nonempty (L ≃ₗ⁅K⁆ (Dim2.Abelian K)) ∨ Nonempty (L ≃ₗ⁅K⁆ (Affine K)) 

theorem Dim3.classification (h : Module.finrank K L = 3) (hs : LieAlgebra.IsSolvable L) :
    Nonempty (L ≃ₗ⁅K⁆ (Dim3.Abelian K)) ∨
    Nonempty (L ≃ₗ⁅K⁆ (Heisenberg K)) ∨
    Nonempty (L ≃ₗ⁅K⁆ (AffinePlusAbelian K)) ∨
    Nonempty (L ≃ₗ⁅K⁆ (Hyperbolic K)) ∨
    (∃ α, α ≠ 0 ∧ Nonempty (L ≃ₗ⁅K⁆ (Family K α 0))) ∨
    (∃ α, α ≠ 0 ∧ Nonempty (L ≃ₗ⁅K⁆ (Family K α 1))) 
\end{lstlisting}
Recall from Table \ref{tab:not} the notation used in Lean code, in comparison to Theorem \ref{thm:main}. Notice that the \lstinline{∨} in the statements is not an exclusive or.  The fact that a Lie algebra is isomorphic to only one Lie algebra in Theorem \ref{thm:main} is the subject of the next subsection.

The precise form of these isomorphisms is immaterial, as it depends on a (noncomputable) choice of basis anyway. Therefore we chose to formalize the classifications with \lstinline{theorem}, which always declares a \lstinline{Prop}-valued term, instead of declaring a term of \lstinline{LieEquiv} for each case. The mere \emph{existence} of a term of type \lstinline{T} can be expressed using the proposition \lstinline{Nonempty T : Prop}, and through the principle of \emph{proof irrelevance} the term is not remembered by Lean.

The insightful reader will notice that the hypothesis of \emph{solvability} is only present in dimension three. In fact, our theorems include the classification of \emph{all} Lie algebras of dimension $\leq 2$ over a field $\K$ -- as very well known, any such Lie algebra is solvable. We obtain this result in Lean as a corollary of our classification.

For dimension one, this is immediate from the instance in mathlib \lstinline{LieAlgebra.ofAbelianIsSolvable}. 
For dimension two, the proof follows from the latter together with the fact that any Lie algebra having a solvable commutator is solvable, which we formalized as \lstinline{LieAlgebra.solvable_of_commutator_solvable}.

In order to prove the classification theorems, given a Lie algebra $L$, we need to provide a Lie algebra isomorphism (\lstinline{LieEquiv}). For abelian Lie algebras we may choose any basis and use the mathlib defition \lstinline{Basis.equivFun}, giving a linear isomorphism between a module over $\K$ with a finite basis and the module of functions from its basis to $\K$.

For the non-abelian Lie algebras we construct a basis satisfying special bracket relations. In dimension two, a unique lemma suffices:
\begin{lstlisting}
lemma Dim2.abelian_or_basis (h : Module.finrank K L = 2) : 
    IsLieAbelian L ∨ (∃ B : Basis (Fin 2) K L, ⁅B 0, B 1⁆ = B 1) 
\end{lstlisting}
The use of bases in our approach is motivated by the fact that Lie theorists usually define Lie algebras in a concise way by giving the non-zero Lie bracket relations between the basis elements. We believe that this way of definition could be a viable way to go forward with in mathlib, possibly together with an automated verification of the Jacobi identity.

With this lemma, the proof of \lstinline{Dim2.classification} goes as follows:
\begin{lstlisting}
theorem Dim2.classification (h : Module.finrank K L = 2) :
    Nonempty (L ≃ₗ⁅K⁆ (Abelian K)) ∨ Nonempty (L ≃ₗ⁅K⁆ (Affine K)) := by
  obtain (a | ⟨B, pfB⟩) := Dim2.abelianorbasis h
  · left
    /- choose any linear isomorphism f, the bracket condition is then trivial -/
    [...]
    exact ⟨{f with [...]}⟩
  · right
    exact ⟨{fun l => ![B.repr l 0, B.repr l 1] with [...]}⟩
\end{lstlisting}
In the left (abelian) case, we choose some linear isomorphism using existing mathlib definitions, and then show that it (trivially) preserves the bracket; in the right case, we explicitly define the isomorphism by sending an element \lstinline{x : L} to a vector in \lstinline{Fin 2 → K} using \lstinline{B.repr}, which gives the coordinates of \lstinline{x} in the basis \lstinline{B} obtained from the lemma above. 

The 3-dimensional case is more involved, since now the commutator ideal $[L,L]$ may have dimension 0, 1 or 2. We provide lemmas with bases in each case.

The 0-dimensional commutator case corresponds to the abelian one. When the commutator is of dimension one, two cases need to be considered; for this part we followed mostly the classification in \cite{Jac}. Under the assumption that the center $\mz(L)$ is contained in the commutator $[L,L]$, we show that \lstinline{∃ B : Basis (Fin 3) K L, ⁅B 1, B 2⁆ = B 0 ∧ ⁅B 0, B 1⁆ = 0 ∧ ⁅B 0, B 2⁆ = 0}. On the contrary, if the center is not contained in the commutator we prove \lstinline{∃ B : Basis (Fin 3) K L, ⁅B 0, B 1⁆ = 0 ∧ ⁅B 0, B 2⁆ = 0 ∧ ⁅B 1, B 2⁆ = B 1}.  These two cases correspond to \lstinline{Heisenberg} and \lstinline{AffinePlusAbelian}, respectively. 
Actually, we implement these as \emph{iff} statements, showing that these bases characterize the relation of the commutator and the center when the commutator is 1-dimensional.

The case of commutator of dimension 2 leads to three possible forms of the bracket; for this part, we were guided by the classification in \cite{DeGraaf:2005}. Here we have
\begin{lstlisting}
lemma Dim3.case2 : Module.finrank K L = 3 ∧ Module.finrank K (commutator K L) = 2 ↔
    (∃ B : Basis (Fin 3) K L, ⁅B 0, B 1⁆ =  B 1 ∧ ⁅B 0, B 2⁆ = B 2 ∧ ⁅B 1, B 2⁆ = 0) ∨ 
    (∃ B : Basis (Fin 3) K L, ⁅B 0, B 1⁆ = B 2 ∧ ⁅B 1, B 2⁆ = 0 ∧
        (∃ α : K, α ≠ 0 ∧ ⁅B 0, B 2⁆ = α • B 1)) ∨ 
    (∃ B : Basis (Fin 3) K L, ⁅B 0, B 1⁆ = B 2 ∧ ⁅B 1, B 2⁆ = 0 ∧
        (∃ α : K, α ≠ 0 ∧ ⁅B 0, B 2⁆ = α • B 1 + B 2))
\end{lstlisting}
The three cases above will correspond to, respectively, the Lie algebras \lstinline{Hyperbolic K}, \lstinline{Family K α 0} and \lstinline{Family K α 1}.

Statements involving bases are the core of the proof of the classification theorem. Once they are established, we proceed as in dimension two, where we define, for each case, an explicit map using \lstinline{B.repr} and show it is a \lstinline{LieEquiv}.
The proof of these key lemmas for $\dim[L,L]=2$ motivated most of the general results that were formalized as part of this project. We shall very briefly describe the proof of the lemmas in a general way, since each has their particularities. We start by setting a basis of the commutator ideal and then extend it to a basis of the Lie algebra. This is done because, as can be seen, the Lie bracket of some basic elements span the commutator, i.e.~the extended bases are \emph{adapted} to the commutator. Finally, we modify this set that a basis and satisfies the required relations. 

The procedure of extending a basis is standard in linear algebra. However, in Lean there are several steps to be dealt with before obtaining a basis of the whole space; the subtleties and some of the auxiliary lemmas used are explained in Section~\ref{sec:linalg}.

\subsection{Non-redundancy of the classification}

Here we describe the theorems and tools that allow to show that the Lie algebras in Theorem \ref{thm:main} are pairwise non-isomorphic.  As we mentioned before, we can restrict the study to Lie algebras of the same dimension, since different dimension Lie algebras cannot be isomorphic. 

In dimension one, there is only one representative, so there is nothing to prove. For dimension two, we proved directly:
\begin{lstlisting}
theorem Dim2.not_iso : IsEmpty ((Affine K) ≃ₗ⁅K⁆ Abelian K)
\end{lstlisting}

For dimension 3, we formalized the non-redundancy in several steps. First, we show that if two Lie algebras are isomorphic then their commutator ideals have equal dimension -- this is actually for any dimension. This is done in \lstinline{LieAlgebra.dim_commutator_eq_of_lieEquiv}. 

In order to use this result, we characterize the non-abelian Lie algebras of dimension three appearing in Table \ref{tab:not}. Indeed we prove that such a Lie algebra $L$:
\begin{itemize}
    \item is isomorphic to $\K^3$ if and only if has 0-dimensional commutator (we use a result valid for any dimension, \lstinline{LieAlgebra.abelian_iff_dim_comm_zero}).
    \item is isomorphic to $\heis_3(\K)$ if and only if has 1-dimensional center and the center is contained in the commutator (\lstinline{Dim3.heisenberg_iff}).
    \item is isomorphic to $\aff(\K)\oplus\K$ if and only if has 1-dimensional center and the center is not contained in the commutator (\lstinline{Dim3.affinePlusAbelian_iff}).
    \item has commutator of dimension two if and only if has a basis satisfying particular bracket relations (\lstinline{Dim3.case2}). We further split this into $\hyp_3(\K)$ and $F_{\alpha,\beta}$ (\lstinline{Dim3.hyperbolic_iff} and \lstinline{Dim3.family_iff}).
\end{itemize}
To conclude that \lstinline{Heisenberg K} is not isomorphic to \lstinline{AffinePlusAbelian K}, we use the characterization above and show that any Lie algebra isomorphism (\lstinline{LieEquiv}) preserves the property of the center being contained in the commutator. Indeed, a Lie algebra satisfying this condition is called \emph{2-step nilpotent}.

The non-isomorphisms of the Lie algebras within the last class requires a bit more of effort. First, we show that \lstinline{Hyperbolic K} cannot be isomorphic to any Lie algebra in \lstinline{Family K α β}:
\begin{lstlisting}
theorem Dim3.Family.not_iso_hyperbolic {α β : K} (hα : α ≠ 0) :
    IsEmpty (Family K α β ≃ₗ⁅K⁆ Hyperbolic K)
\end{lstlisting}
The proof of this theorem uses a more subtle invariant than the 2-step condition. Indeed, we use the fact that there is an element $e\in\hyp_3(\K)$ whose bracket $\ad(e):=[e,\cdot\,]$ acts as the identity on the commutator. More precisely, as mentioned above,  $\hyp_3(\K)$ admits a basis $\mathcal B=\{b_0, b_1, b_2\}$ satisfying the first line of the lemma \lstinline{Dim3.case2}. With that it is possible to prove in Lean that the commutator of this Lie algebra is spanned by $\{b_1, b_2\}$. In addition, the well defined endomorphism of the commutator defined as the restriction of $\ad(b_0)$ is the identity map. If a Lie algebra is isomorphic to $\hyp_3(\K)$, then it must contain an element such that the induced endomorphism is also the identity. The proof of \lstinline{Dim3.Family.not_iso_hyperbolic} shows that \lstinline{Family K α β} does not possess such an element.

As a last step, we provide criteria to decide when two members of \lstinline{Family K α β} are isomorphic. This criteria completes the proof of Theorem \ref{thm:main} (see notation in Table \ref{tab:not}).

\begin{lstlisting}
theorem Dim3.Family.iso_iff {α α' β β' : K} (hα : α ≠ 0) (hα' : α' ≠ 0) :
  Nonempty ((Family K α β) ≃ₗ⁅K⁆ Family K α' β') ↔ ∃ (γ : Kˣ),  α = γ^2 * α' ∧ β = γ * β' 
\end{lstlisting}
For this proof, we consider a basis $\mathcal B=\{b_0, b_1, b_2\}$ of \lstinline{Family α β} such that the Lie bracket satisfy 
$[b_0, b_1] = b_2, [b_1, b_2] = 0, [b_0, b_2] = \alpha  b_1 + \beta  b_2$. One shows in Lean that the commutator is spanned by $b_1,b_2$. As done for the hyperbolic, we consider the endomorphism of the commutator induced by taking brackets with $b_0$. This endomorphism has a matrix, in the basis $\{b_1,b_2\}$, of trace $\beta$ and determinant $-\alpha$. Therefore, if there is an isomorphism $f: F_{\alpha,\beta}\to F_{\alpha',\beta'}$, then $f$ takes the commutator  to the commutator of the image (\lstinline{LieEquiv.commutator_map}). The image of $b_0$ by the isomorphism induces a map in the commutator which is a multiple $\gamma$ of the conjugation of the map induced by $b_0$ on the domain. Since conjugation preserves trace and determinant, we obtain our result. 

We note that the adjoint maps $x\mapsto \ad(x)=[x,\cdot\,]$ of a Lie algebra are already formalized in mathlib as \lstinline{LieDerivation.ad}. For the proof of the last theorem, we formalized and used in Lean the restriction of these maps to the commutator. In this way we were able to compute trace and determinants of the $2\times 2$ matrix induced and to show that these were conjugate, up to a multiple. 

\section{Further contributions}
\label{sec:contr}

The formalization of the results in Section \ref{sec:classif} required us to implement along the way many auxiliary results about modules, bases, Lie algebras and their substructures. Some are completely new theorems and constructions, other are marginal improvements or adjustments of content already in mathlib. In what follows we give a rough overview over these contributions, including only some selected examples due to the limited scope of this paper.

Many of the following definitions and theorems can and have been stated in high generality, which is one of the design principles of mathlib. For example, vector spaces over fields are just a special case of modules over a ring, but this can be even further generalized to modules over a semiring. This is all encompassed by the typeclass \lstinline{Module} in mathlib. For the simplicity of exposition, we will state the following theorems with vector spaces and Lie algebras over fields. As we will shortly see, from a working mathematician's and user's perspective it also sometimes makes sense to run contrary to this design philosophy by specializing some definitions to particular cases and giving them new names.

\subsection{Linear algebra lemmas}
\label{sec:linalg}

The proofs of the classification theorems that we formalized required explicit calculations in finite-dimensional vector spaces on the level of submodules, linear independent sets of vectors, and bases. In mathlib there exist typeclasses for all of these, but stating even mathematically simple statements can quickly become convoluted. For example, let $V$ be a vector space over a field $K$ -- that is, \lstinline{K V : Type*} and there are instances of the typeclasses \lstinline{Field K}, \lstinline{AddCommGroup V} and \lstinline{Module K V}. Recall the definition of \lstinline{Submodule K V} as a structure type from Section~\ref{sec:prelimlean}. A term \lstinline{p : Submodule K V} corresponds to a vector subspace of $V$. Trivially, this should be itself a vector space over $K$. But in order to equip it with a \lstinline{Module} instance, it first needs to be coerced to a \lstinline{Type}. This is done by the \lstinline{Subtype} structure: we obtain a term \lstinline{↥p : Type*} and a canonical function \lstinline{Subtype.val : ↥p → V}, and mathlib provides us with an instance of \lstinline{Module K ↥p}.

This subtlety of type coercions appears all over the place and adds a bit more ballast to the task of formalization, but is ultimately unavoidable. It explains the need to prove mathematically trivial lemmas such as the following two from our project.\pagebreak
\begin{lstlisting}
variable {K L : Type*} [Field K] [AddCommGroup L] [Module K L]

theorem LinearIndependent.iff_in_submodule {S : Type*} (N : Submodule K L) {f : S → N} :
    LinearIndependent K f ↔ LinearIndependent K (Subtype.val ∘ f: S → L)
\end{lstlisting}

\begin{lstlisting}
theorem Submodule.linearIndependent_from_ambient {S : Type*} (N : Submodule K L) (f : S → L)
         (hs : LinearIndependent K f) (hr : Set.range f ⊆ N) :
    LinearIndependent K (Set.map_into_subtype N f hr)
\end{lstlisting}

An elementary method in linear algebra is to start with a linearly independent set of vectors, or a basis of some subspace, and then extend it to a basis of the whole vector space. Although there already exist constructions in mathlib that help with this, we felt the need to write some additional API, such as the following that extends a linear independent set of size one less than the rank of the module to a basis.
\begin{lstlisting}
noncomputable def mkFinCons_from_linIndep {n : ℕ} {l : Fin n → L}
        (hs : LinearIndependent K l) (ht : Module.finrank K L = n + 1) :
    Basis (Fin (n + 1)) K L
\end{lstlisting}

\subsection{Lemmas about Lie algebras}

As in the preceding section, some of the lemmas we formalized are of a more technical flavor and some are more mathematically meaningful. One such technical lemma, which also exists under the alias \lstinline{LieIdeal.coe_map_of_surjective} in mathlib, states that if a Lie algebra homomorphism $f: L\to L'$ is surjective, then the pushforward of an ideal $I\subseteq L$ (i.e.~the ideal generated by $f(I)$) is just the image $f(I)$ itself. Because ideals are in particular subalgebras, subspaces and subsets, which all correspond to different structure types in mathlib, this statement can be formalized on several levels:
\begin{lstlisting}
variable {K L L': Type*} [Field K] [LieRing L] [LieRing L'] [LieAlgebra K L] [LieAlgebra K L']

theorem LieIdeal.map_eq_image_of_surj {f : L →ₗ⁅K⁆ L'}
        (h : Function.Surjective f) (I : LieIdeal K L) :
    (LieIdeal.map f I).toSubmodule = Submodule.map f.toLinearMap I

theorem LieIdeal.map_eq_image_of_surj' {f : L →ₗ⁅K⁆ L'}
        (h : Function.Surjective f) (I : LieIdeal K L) :
    (LieIdeal.map f I).toLieSubalgebra = LieSubalgebra.map f I

theorem LieIdeal.map_eq_image_of_surj'' {f : L →ₗ⁅K⁆ L'}
        (h : Function.Surjective f) (I : LieIdeal K L) :
    LieIdeal.map f I = f '' I
\end{lstlisting}
Note that \lstinline{LieIdeal.map_eq_image_of_surj} already exists in mathlib  as \lstinline{LieIdeal.coe_map_of_surjective}, but having the above alternative versions was very useful to us. For example, it allowed us to construct, starting from an isomorphism $e: L\to L'$ between two Lie algebras, an isomorphism between any ideal $I\subseteq L$ and its image.
\begin{lstlisting}

def LieEquiv.ofIdeals (e : L ≃ₗ⁅K⁆ L') (I : LieIdeal K L) (I' : LieIdeal K L')
        (h : LieIdeal.map e.toLieHom I = I') :
    I ≃ₗ⁅K⁆ I'
\end{lstlisting}
This is analogous to the mathlib theorems \lstinline{LinearEquiv.ofSubmodules} and \lstinline{LieEquiv.ofSubalgebras}.

To present another example of coercion between types, consider the derivations of an abelian Lie algebra $L$. Since the bracket on $L$ is trivial, the derivation condition is vacuous, so every linear endomorphism of $L$ is indeed a derivation. This trivial fact is formalized in the following isomorphism of Lie algebras.\pagebreak
\begin{lstlisting}
def Abelian.DerivationOfLinearMap (K L : Type*) [CommRing K] [LieRing L] 
        [LieAlgebra K L] [IsLieAbelian L] :
    End K L ≃ₗ⁅K⁆ LieDerivation K L L
\end{lstlisting}

Returning to ideals in a Lie algebra, we note the commutator and center ideals were instrumental in the proof of the classification theorems. Even though the commutator ideal is just the first term of the derived series (which is already implemented in mathlib), we deem it worthy of having its own name in Lean. This makes theorem statements and proofs easier to read and think about without introducing additional abstraction. The theorem \lstinline{LieAlgebra.commutator_eq_span} in turn is vital for the calculational parts of the proofs.
\begin{lstlisting}
abbrev LieAlgebra.commutator : LieIdeal K L := LieAlgebra.derivedSeries K L 1
theorem LieAlgebra.commutator_eq_span :
    (LieAlgebra.commutator K L).toSubmodule =
        Submodule.span K {x : L | ∃ (y : L) (z : L), ⁅y, z⁆ = x}
\end{lstlisting}
We also needed the fact that commutator and center are preserved under isomorphisms of Lie algebras:
\begin{lstlisting}
def LieEquiv.commutator_equiv (e : L ≃ₗ⁅K⁆ L') : LieAlgebra.commutator K L ≃ₗ⁅K⁆ LieAlgebra.commutator K L'
theorem LieEquiv.center_map (e : L ≃ₗ⁅K⁆ L') : LieIdeal.map e (LieAlgebra.center K L) = LieAlgebra.center K L'
\end{lstlisting}
In the first case, the Lie algebra structure of the commutator itself was important, hence the need to define an isomorphism. In the second case, the center is always itself abelian, so there is no information in the Lie algebra structure.

As a concrete interaction between linear algebra concepts and the Lie bracket, there is the elementary consequence of the skew-symmetry of the bracket that two vectors $x,y\in L$ with $[x,y]\neq0$ are necessarily linearly independent.
\begin{lstlisting}
lemma LieAlgebra.linearIndependent_of_bracket_ne_zero (X Y : L) (hXY : ⁅X, Y⁆ ≠ 0) :
    LinearIndependent K ![X, Y]
\end{lstlisting}

An important fact of a more general interest is that a Lie algebra $L$ is solvable by taking an ideal $I\subseteq L$ and showing that both $I$ and the quotient Lie algebra $L/I$ are solvable.
\begin{lstlisting}
theorem LieAlgebra.solvable_of_ideal_and_quot_solvable {I : LieIdeal R L}
        (quotsol : LieAlgebra.IsSolvable (L ⧸ I)) (Isol : LieAlgebra.IsSolvable I) :
    LieAlgebra.IsSolvable L
\end{lstlisting}

\subsection{The semidirect product}
\label{sec:semidirect}

We turn to a central construction in Lie theory that allows us to define many Lie algebras of interest without resorting to a basis or coordinates. Let $(L,\br_L)$ and $(J,\br_J)$ be two Lie algebras (over the same field) such that $L$ acts on $J$ in the sense of a representation of Lie algebras, that is, we fix a Lie algebra homomorphism $\varphi: L\to\Der J$. Then the \emph{semidirect product} $L\ltimes_\varphi J$ is defined as the direct product of vector spaces $L\times J$ together with a Lie bracket given by
\[[(x_1,x_2),(y_1,y_2)]:=([x_1,y_1]_L,[x_2,y_2]_J+\varphi(x_1)y_2-\varphi(y_1)x_2).\]
This is analogous to group representations and the semidirect product in group theory, which is already formalized in mathlib:
\begin{lstlisting}
structure SemidirectProduct (N G : Type*) [Group N] [Group G] (φ : G →* MulAut N) where
    left : N
    right : G
instance {N G : Type*} [Group N] [Group G] {φ : G →* MulAut N} : 
    Mul (SemidirectProduct N G φ) where
    mul a b := ⟨a.1 * φ a.2 b.1, a.2 * b.2⟩
\end{lstlisting}
In our definition of semidirect product of Lie algebras \lstinline{L J : Type*}, we decided to directly use the product type \lstinline{L × J}, so we can use its existing \lstinline{Module} instance.
\begin{lstlisting}
def LieSemidirectProduct {K : Type*} (L J : Type*) [Field K] [LieRing L] [LieRing J]
        [LieAlgebra K L] [LieAlgebra K J] (_ : L →ₗ⁅K⁆ LieDerivation K J J) := 
    L × J
\end{lstlisting}
We also set the shorthand notation \lstinline{L ⋉[φ] J} for 
\lstinline{LieSemidirectProduct L J φ}.
\begin{lstlisting}
variable {K L J : Type*} [Field K] [LieRing L] [LieRing J] [LieAlgebra K L] [LieAlgebra K J]
    {φ : L →ₗ⁅K⁆ LieDerivation K J J}
    
instance : AddCommGroup (L ⋉[φ] J) := (inferInstance : AddCommGroup (L × J))
instance : Module K (L ⋉[φ] J) := (inferInstance : Module K (L × J))
instance : Bracket (L ⋉[φ] J) (L ⋉[φ] J) where
    bracket := fun a b ↦ ⟨⁅a.1, b.1⁆, φ a.1 b.2 - φ b.1 a.2 + ⁅a.2, b.2⁆⟩
\end{lstlisting}
Along the way we wrote extensionality and simplification lemmas, as well as the original Lie algebras $L$ and $J$ as subalgebras of $L\ltimes_{\varphi} J$. In fact, $J$ is even an ideal, but contrary to the direct product, $L$ is generally not.
\begin{lstlisting}
def LieSemidirectProduct.leftSubalgebra : LieSubalgebra K (L ⋉[φ] J) := LieHom.range inl
def LieSemidirectProduct.rightIdeal : LieIdeal K (L ⋉[φ] J) := LieHom.ker fst
\end{lstlisting}
Finally, for the trivial representation $\varphi=0$, the semidirect product just reduces to the direct product:

\begin{lstlisting}
def Prod.toLieSemidirectProduct : (L × J) ≃ₗ⁅K⁆ L ⋉[(0 : L →ₗ⁅K⁆ LieDerivation K J J)] J
\end{lstlisting}

The Lie algebras appearing in Theorem~\ref{thm:main} can in fact all be written as direct/semidirect products of smaller pieces. We defined the corresponding isomorphisms, and moreover also formalized two general series of Lie algebras, $\aff(\K^n)=\End(\K^n)\ltimes\K^n$ and $\hyp_n(\K)=\K\ltimes\K^{n-1}$.

\subsection{Almost abelian Lie algebras}
\label{sec:alab}
An interesting class of Lie algebras is that of \emph{almost abelian} ones. We call a Lie algebra $L$ almost abelian if $L$ contains an abelian ideal $I$ of codimension one, that is $\dim L/I=1$. In this case all of the Lie bracket information can be encoded in a single linear endomorphism of $I$, which makes almost abelian Lie algebras very pleasant study objects. We formalized the definition (at least for finite-dimensional Lie algebras):
\begin{lstlisting}
variable (K L : Type*) [Field K] [LieRing L] [LieAlgebra K L]
def LieAlgebra.IsAlmostAbelian : Prop :=
    ∃ I : LieIdeal K L, IsLieAbelian I ∧ Module.finrank K L = Module.finrank K I + 1
\end{lstlisting}
A semidirect product of the form $\K\ltimes_\varphi L$, where $L$ is abelian, is clearly almost abelian.
\begin{lstlisting}
theorem LieSemidirectProduct.isAlmostAbelian {φ : K →ₗ⁅K⁆ LieDerivation K L L}
        [IsLieAbelian L] :
    LieAlgebra.IsAlmostAbelian K (K ⋉[φ] L)
\end{lstlisting}
Together with the formalized isomorphisms between the low-dimensional Lie algebras in the classification and semidirect products of this form, one may conclude that every solvable Lie algebra of dimension at most three over a field is almost abelian.

In the converse direction, we also formalized the theorem that if a (not necessarily abelian) Lie algebra $L$ has a codimension one ideal $I$, it is isomorphic to a semidirect product $\K\ltimes_\varphi I$.
\begin{lstlisting}
theorem LieAlgebra.semidirectProduct_of_codim_one_ideal (I : LieIdeal K L)
        (hdim : Module.finrank K L = Module.finrank K I + 1) :
    ∃ φ : K →ₗ⁅K⁆ LieDerivation K I I, Nonempty (L ≃ₗ⁅K⁆ K ⋉[φ] I)
\end{lstlisting}
Expressing the Lie algebras in Theorem \ref{thm:main} as the semidirect products mentioned above allowed us to prove in Lean the very well known fact that every 3-dimensional Lie algebra is almost abelian. 

\section{Development}\label{sec:dev}

Our development is based on Lean 4 and mathlib4, both on version \lstinline{v4.19.0}, and can be accessed on \href{https://github.com/LieLean/LowDimSolvClassification}{\lstinline|https://github.com/LieLean/LowDimSolvClassification|}.

We have decided to work on top of Lie algebras as currently defined in mathlib. As a result of our restriction to low dimensional Lie algebras, we have benefited from the results around finite bases for vector spaces. We have made use of classical axioms through the development, either in direct form or by using mathlib results that make use of these axioms. In particular, we found ourselves needing to choose elements for extending existing linear independent sets and bases, often leading to \lstinline{noncomputable} definitions. One such example is \lstinline{LinearIndependent.extend_fin_succ_fun}, where we use \lstinline{Classical.choose}.

The tactic \lstinline{module} was useful in certain cases for discharging proofs of Lie algebra equalities, although in some cases we found ourselves needing to apply commutation to Lie atoms of the form \lstinline{⁅x, y⁆} before applying the tactic. To get around this restriction, we created a modified version of \lstinline{module}, which we call \lstinline{simplify_lie}, and adds additional support to atoms of the form \lstinline{⁅x,y⁆}. This tactic will perform rewriting of Lie equalities (\lstinline{lie_add}, \lstinline{add_lie}, \lstinline{lie_smul}, \lstinline{smul_lie}, \lstinline{lie_neg}, \lstinline{neg_lie}, \lstinline{sub_lie} and \lstinline{lie_sub}), and then proceed with a \lstinline{match_scalars} modification that will keep lists of items consisting of two forms: either of the form \lstinline{⁅x,y⁆}, or of the regular form \lstinline{x}, where \lstinline{x} and \lstinline{y} represent expressions that are considered atoms. When forming the equations to be solved, this modified \lstinline{match_scalars} will match \lstinline{⁅x, y⁆} to both \lstinline{⁅x, y⁆} and \lstinline{⁅y, x⁆} (and keep the fact that it goes in negative form) by using a modified version of \lstinline{AtomM}. In practice, however, we did not find this modification very useful, and we hoped to define an improved tactic that could handle other Lie equations (such as Lie associativity) by finding appropriate normal forms.
After the first draft version of this article, we found an open PR~\cite{tactic-liering} on mathlib which introduces a tactic for Lie algebras. It is based on Lyndon words for building the normal forms, and the tactic seems to handle all the equations we wanted to handle. We plan to make use of this tactic in the future.

We have extracted the theorems and lemmas that we believe could be useful more generally into separate files, and plan to submit a pull request to mathlib4 in the future. 
The following is a short description for the main files in the development:
\begin{itemize}
\item \verb|Classification1.lean|, \verb|Classification2.lean|, \verb|Classification3.lean|: these are the main files introducing the classification theorems for each dimension.
\item \verb|GeneralResults.lean|: general results used in the classification that we aim to contribute to mathlib.
\item \verb|LemmasDim3.lean|: particular lemmas needed in the proof, related to dimension 3.
\item \verb|InstancesConstructions.lean|, \verb|InstancesLowDim.lean|: definitions of Lie algebras (general constructions and low dimensional instances used in classification).
\item \verb|Semidirect.lean|: introduces the semidirect product of Lie algebras.
\item \verb|QuotientSolvable.lean|: the proof that a Lie algebra is solvable if a quotient by some solvable ideal is solvable.  
\item \verb|Tactics.lean|: implements the \lstinline{simplify_lie} tactic based on Macbeth's \lstinline{module} and Carneiro's \lstinline{AtomM}.
\end{itemize}

\section{Future work}\label{sec:future}

The formalization of Theorem~\ref{thm:main} led to the mechanization of several well-known results in Lie theory, many of which could potentially be added to mathlib. However, there is still a lot of work to be done.

Looking ahead, we see two promising directions for future work. One is to extend the classification to solvable Lie algebras of dimension $\geq 4$. Whenever possible, we have stated theorems and lemmas in a way that applies to arbitrary finite dimension. However, the main classification proofs in this paper rely on ad hoc analysis of specific bases by dimension. So far, moving to higher dimensions would require new adapted bases and a case-by-case analysis. It would be interesting to find proof strategies that allow reusing lower-dimensional results in higher-dimensional cases. We believe that the implementation of semidirect products may be helpful in this regard. In fact, any solvable Lie algebra of dimension $n$ is a semidirect product of the field of scalars and a solvable Lie algebra of dimension $n-1$.

Another direction is to complete the picture in dimension three by formalising a classification of simple Lie algebras. In characteristic zero, this is definitely doable. However, in positive characteristic, the choice of the classification theorem to be formalized must be made carefully, as different theorems may yield multiple non-isomorphic classes depending on the characteristic \cite{FW14,Jac,Strade}.

\bibliographystyle{alpha}
\bibliography{main}

\medskip

\noindent {\bf Viviana del Barco, Gustavo Infanti} and {\bf Paul Schwahn:}  Instituto de Matem\'atica, Estat\'istica e Computa\c{c}\~ao Cient\'ifica, Universidade Estadual de Campinas, Rua Sergio Buarque de Holanda, 651, Cidade Universitaria Zeferino Vaz, 13083-859, Campinas, S\~ao Paulo, Brazil.

\noindent {\it Email address:} {\tt delbarc@unicamp.br, g246337@dac.unicamp.br, schwahn@unicamp.br }
\medskip

\noindent{\bf Exequiel Rivas:} Dept.~of Software Science,  Tallinn University of Technology, Akadeemia tee 21, 12618 Tallinn, Estonia.

\noindent {\it Email address:} {\tt exequiel.rivas@taltech.ee}

\end{document}